\documentclass{article}
\usepackage{amssymb}

\usepackage[dvips]{epsfig}
\newcommand{\sta}{{\rm sta}}
\newcommand{\ext}{{\rm ext}}

\newcommand{\barx}{\bar{x}}
\newcommand{\bary}{\bar{y}}

\newcommand{\real}{{\mathbb R}} %\newcommand{\real}{{\bf R}}
\newcommand{\half}{\frac{1}{2}}

\newcommand{\eba}{\begin{array}}
\newcommand{\eea}{\end{array}}
\newcommand{\ebe}{\begin{eqnarray}}%[section]
\newcommand{\eee}{\end{eqnarray}}%[section]
\newcommand{\eb}{\begin{equation}}%[section]
\newcommand{\ee}{\end{equation}}%[section]

\newcommand{\calP}{{\cal{P}}}

\newcommand{\bx}{{\bf x}}

\newcommand{\calS}{{\cal S}}

\newcommand{\alp}{{\alpha}}

\newcommand{ \Lam}{{\Lambda}}

\newcommand{ \lam}{{\lambda}}

\newtheorem{remark}{Remark}

\newtheorem{thm}{Theorem}

\begin{document}

\begin{center}
{\Large \textbf{Complete solutions to a Class of  \\
 8$^{th}$ Order Polynomial Optimization Problems}
\vspace{0.4cm}\\[0pt]}
{\textbf{Timothy K. Gao   }}\\
  College of Engineering, Cornell University, Ithaca, NY 14853, USA. E-mail: tkg24@cornell.edu\\

\end{center}

\date{}
%\maketitle
\begin{abstract}
This paper presents a new class of  8$^{th}$ order  canonical polynomials in $\real^n$. Based on the sequential canonical dual transformation,
all extrema are obtained.
The method can be used to solve the associated
7$^{th}$ order nonlinear algebraic equations.
Optimality  conditions are provided to find both global
minimizers and local extrema. Applications are illustrated by examples.
\end{abstract}

{\bf Key Words:} Canonical polynomials; triality theory;
nonlinear algebraic equations; canonical duality.

\section{Primal Problem and Motivation}
This paper intends to solve the following  8$^{th}$ order  polynomial minimization problem
 \eb
(\calP_n)\;\;\;\;\min \left\{P_n({\bf x}) =  V(\bx)  - {\bf x}^T {\bf h}
|\; {\bf x} \in \real ^n \right\}, \label{eq-8polyn}
\ee
where 
\begin{eqnarray}
 V(\bx) &=& \half a_2(y_2(y_1({\bf x})))^2+ b_2 y_2(y_1({\bf x}))+c_2, \\
y_2 (y_1) &=& \half a_1 (y_1)^2+b_1 y_1  + c_1,\\
y_1({\bf x}) &=& \half a_0\|{\bf x}\|^2 + {\bf x}^T {\bf b_0}+c_0,
\end{eqnarray}
and $a_0, c_0, a_i , b_i,  c_i \; ( i = 1,2)$  are given constants, $ {\bf h}, {\bf b_0} \in \real^n$ are given vectors.
In this paper, we
assume that $a_i >  0  \; ( i =  0,1, 2)$ because $(\calP_n)$ is a minimization problem.

The global optimization problem $(\calP_n)$ appears extensively in engineering physics and materials science:
 examples include phase transitions of solids \cite{gao-ogden07}, shape-memory alloys \cite{stanly}, and superconductivity
 \cite{gao-yu}.
 From a systems theory point of view, if the vector ${\bf h} \in \real^n$ is considered as a source  (input) variable, say the external force in structural mechanics, 
 then the linear term $\bx^T {\bf h}$ can be viewed as the ``external energy"  and  the polynomial $V(\bx)$ 
 as the ``internal (or free) energy". Clearly, if ${\bf b_0} = {\bf 0} \in \real^n$, then 
 $V(\bx)$ is the so-called objective function \cite{gao-book00}, i.e.,
 $V(Q \bx) = V(\bx)$ for all $\bx \in \real^n$ and all $Q \in \real^{n\times n}$ such that $Q^{-1} = Q^T$ and $\det Q = 1$. 
 Moreover, if $a_2 = b_1 = c_1 =0$ and $c_0 = - \lam < 0$, then $V(\bx)$ is the well-known {\em  double-well potential}
 \[
W(\bx) = \half \alp \left( \half a_0 \| \bx \|^2 - \lam \right)^2 , \;\; \alp = b_2 a_1 > 0,
\]
which was first proposed by van der Waals in thermodynamics and which appears in many branches of science. In theoretical physics, it provides one of the simplest models of the unified field theory.  In ferroelectric and ferromagnetic systems, the critical
phenomena and the phase transition  are modeled by this double well potential \cite{bodineau,brauman,kaski}. 
It can also be found in the theory of dislocations in metals, the theory of Josephson junctions, nonconvex dynamics, computational biology, and systems of network communication  \cite{gao-amma03,ruan-gao-jiao,zgy}. 
Particularly, the graph of $W(\bx)$ in $\real^2$ is the so-called Mexican hat potential, which has been used to model the Higgs Mechanism in quantum mechanics and is also referred to as the {\em sombrero potential} in cosmic strings \cite{Copeland-Kibble}.
 In order to model high-order phase transitions, a  high order van der Waals-type polynomial has been proposed in \cite{gao-book00,gao-jogo06}. Because the polynomial $V(\bx)$ proposed in this paper has more coefficients than the van der Waals-type polynomials, it can be used for modeling more complicated phenomena. 
 However, due to its nonconvexity, the problem $(\calP_n)$ could have multiple local minimizers, and finding the globally optimal solution by traditional direct methods is considered a fundamentally challenging task in global optimization.

Canonical duality theory developed in \cite{gao-book00} is a potentially useful methodology which can be used to solve a relatively large class of nonlinear algebraic and differential equations \cite{gao-jogo06, gao-ogden07}. The associated triality theory can be used to identify both local extrema and global minimizer (see \cite{gao-wu-jimo}). The purpose of this paper is to generalize this result to solve the global optimization problem $(\calP_n)$. In this next section, we first  show that the $n$-dimensional nonconvex polynomial minimization problem can be reformulated by the canonical duality theory into a one-dimensional canonical dual problem, which can be solved easily. Therefore, an analytical solution to $(\calP_n)$ is presented in Sections 2, 3, and 4. We then discuss the special case of $n=1$ in Section 5 and show that the one-dimensional 8$^{th}$-order polynomial optimization problem can be solved completely by the canonical duality theory to obtain all critical points. The extremality behaviors of all these critical solutions can be identified by a generalized triality theory. Applications are illustrated in Section 6. The paper ends with some conclusion remarks.

\section{$P_n({\bf x})$ and Its Canonical Dual}
From the pattern of the 8$^{th}$ order polynomial defined in (\ref{eq-8polyn}), we introduce
a geometrical operator
\eb
y_1=\Lam_1({\bf x})=\half  a_0 \|{\bf x}\|^2 + {\bf b_0}^T {\bf x} + c_0,
\ee
so $y_2({\bf x})= U_1(\Lam_1({\bf x}))$ and $U_1 (y_1)= \half a_1 y_1^2+b_1 y_1 + c_1$.
The dual variable  of $y_1$ is defined as
\eb\label{s1}
s_1 = \frac{dU_1(y_1)}{dy_1} = a_1 y_1 +b_1.
\ee
Thus $(y_1, s_1)$ form a \textit{canonical duality pair} \cite{gao-book00} and
the Legendre conjugate $U_1^*$ can be uniquely defined by
\begin{eqnarray}\label{trans}
U_1^* (s_1) &=& \sta\{y_1  s_1 - U_1(y_1)|\; y_1 \in \real\}\nonumber\\
&= & \frac{(s_1 - b_1 ) s_1}{a_1} - \left[\frac{(s_1 - b_1)^2}{2a_1} + \frac{b_1(s_1 - b_1)}{a_1} +  c_1\right]\nonumber\\
&=& \frac{(s_1 - b_1)^2}{2 a_1} - c_1.
\end{eqnarray}
It is easy to verify that
\[
s_1 = \frac{dU_1(y_1)}{dy_1}\Leftrightarrow
y_1 = \frac{dU_1^*(s_1)}{ds_1}\Leftrightarrow
 U_1(y_1) +  U_1^*(s_1)=y_1 s_1.
\]

We introduce a second geometrical operator
\eb\label{qua}
y_2= \Lam_2(y_1)= \half a_1 y_1 ^2 + b_1 y_1  + c_1
\ee
and let
\[
V({\bf x}) = U_2\left( \Lam_2 \left( y_1( {\bf x}) \right) \right),
\]
where $U_2(y_2)= \half a_2 y_2^2+ b_2 y_2+c_2$.
The duality relation between $y_2$ and $s_2$ is
\eb\label{s2}
s_2 = \frac{dU_2 (y_2)}{dy_2} = a_2 y_2 +b_2 .
\ee
Also, $(y_2, s_2)$  form a \textit{canonical duality pair} \cite{gao-book00} and
the Legendre conjugate $U_2^*$ can be uniquely defined by
\begin{eqnarray}\label{transb}
U_2^* (s_2) &=& \sta\{y_2  s_2 - U_2(y_2)| \;y_2 \in \real\}\nonumber\\
&=& \frac{(s_2 - b_2)^2}{2 a_2} - c_2.
\end{eqnarray}
In the same way as before, we can verify that
\[
s_2 = \frac{dU_2(y_2)}{dy_2}\Leftrightarrow
y_2 = \frac{dU_2^* (s_2)}{ds_2}\Leftrightarrow
U_2(y_2) +  U_2^*(s_2)= s_2 y_2.
\]

Since we can replace $V({\bf x}) = U_2(\Lam_2(y_1({\bf x}))) $ with $\Lam_2(y_1({\bf x}))s_2 - U_2^*(s_2)$,
the generalized total complementary function  \cite{gao-jogo06} $\hat{\Xi}: \real^n \times \real
\rightarrow \real$ can be defined as
\[
\hat{\Xi}({\bf x}, s_2)=\Lam_2(y_1({\bf x}))s_2 - U_2^*(s_2)- {\bf h}^T {\bf x}.
\]
Similarly,  $\Lam_2(y_1({\bf x}))=y_2({\bf x}) = U_1(\Lam_1({\bf x}))$  can be replaced by
$\Lam_1( {\bf x} ) s_1 - U_1^*(s_1)$, so the generalized total complementary function  $\hat{\Xi}({\bf x}, s_2)$ can be reformulated as
\begin{eqnarray}
\Xi({\bf x}, s_1 ,s_2)&=& [\Lam_1({\bf x} ) s_1 - U_1^*(s_1)]s_2 - U_2^*(s_2) - {\bf h}^T {\bf x} \label{npria}\nonumber\\
&=& \left[ \left(\half a_0 \|{\bf x}\|^2 + {\bf b_0}^T {\bf x} +c_0 \right) s_1 - U_1^*(s_1) \right ]s_2 - U_2^*(s_2) - {\bf h}^T {\bf x}. \label{nprib}
\end{eqnarray}
By the criticality condition
$\nabla_{\bf x} \Xi({\bf x}, s_1, s_2)={\bf 0}$, it is easy to obtain
\begin{equation}\label{xrelss}
{\bf x} = {\bf x}(s_1, s_2)= \frac{1}{a_0}\left(\frac{\bf h}{s_1 s_2} - {\bf b_0}\right).
\end{equation}
Thus,  the canonical dual function $P^d: \real \times \real \rightarrow \real$ can be defined as
\begin{eqnarray}
P^d(s_1, s_2) &=& \sta \{ \Xi ({\bf x}, s_1, s_2)|\; {\bf x} \in \real^n\} \\
&=& s_1 s_2 \left(c_0 - \frac{1}{2 a_0} \left \| \frac{{\bf h}}{s_1 s_2}- {\bf b_0}\right \|^2  \right)-s_2 U_1^*(s_1) -U_2^*(s_2). \label{Pds1s2}
\end{eqnarray}

Combining  (\ref{s1}), (\ref{qua}), and (\ref{s2}) captures the relation between $s_1$ and $s_2$ given $s_1= a_1 y_1 + b_1$ and $s_2= a_2 y_2 + b_2.$
Therefore we obtain
\begin{eqnarray}\label{rela}
s_2 = s_2(s_1) &= & a_2 \left(\half a_1 \left(\frac{s_1 - b_1 }{a_1}\right)^2
+ b_1 \frac{s_1 - b_1 }{a_1} + c_1\right) +b_2 \nonumber\\
&=&a_2 \left( \frac{s_1^2 - b_1^2}{2a_1} + c_1\right) + b_2 = \frac{a_2}{2 a_1} (s_1^2 - H_3),
\end{eqnarray}
where
\[
H_3 = - \frac{2 a_1 a_2 c_1 + 2 a_1 b_2 - a_2 b_1^2}{a_2}.
\]
To eliminate $s_2$, we let  $\sigma = s_1$ and  $s_2 = \tau(\sigma) =  \frac{a_2}{2 a_1} (\sigma^2 - H_3)$ to convert the total complementary function $\Xi({\bf x}, s_1, s_2)$ to
\eb\label{XiSimp}
\Xi({\bf x}, \sigma)= \Lam_1({\bf x}) \sigma \tau(\sigma) - U_1^*(\sigma) \tau(\sigma) - U_2^*(\tau(\sigma)) - {\bf h}^T {\bf x}.
\ee
Also, (\ref{xrelss}) becomes
\eb\label{xrels}
{\bf x}(\sigma) = {\bf x}(\sigma, \tau(\sigma))= \frac{1}{a_0}\left(\frac{\bf h}{\sigma \tau(\sigma)} - {\bf b_0}\right)
\ee
and the canonical dual function $P^d(s_1, s_2)$ can be converted to
\eb\label{Pds}
P^d(\sigma) =H_4 + \frac{  a_2 (\sigma^2 - H_3)^2}{8 a_1^2} - \frac{(\Phi(\sigma)^2 + H_1)}{a_2 \sigma (\sigma^2 - H_3 )},
\ee
where
\begin{eqnarray}
H_1 &=& \frac{a_1 \|{\bf h}\|^2 }{a_0} \nonumber\\
H_2 &=& \frac{2 a_0 a_1 c_0 + 2 a_0 b_1 - a_1 \|{\bf b_0}\|^2}{2 a_0} \nonumber\\
H_4 &=& \frac{2 a_0 a_2 c_2 + 2 a_2 {\bf b_0}^T{\bf h} - a_0  b_2^2}{2 a_0 a_2}  \nonumber\\
\Phi (\sigma)^2 &=& 2\left[\sigma \tau(\sigma)\right]^2 ( \sigma - H_2) = 2 \left[\frac{a_2 \sigma(\sigma^2 - H_3 )}{2 a_1}\right]^2(\sigma - H_2)\nonumber.
\end{eqnarray}
It can be seen from (\ref{Pds}) that $P^d(\sigma)$ is asymptotic only when ${\bf h} \ne {\bf 0}$.
By the criticality condition
\eb\label{dPds}
\frac{d P^d (\sigma)}{d\sigma}= \frac{a_2 (3 \sigma^2 - H_3)(H_1 - \Phi(\sigma)^2)}{\left[2 a_1\sigma \tau(\sigma)\right]^2}=0,
\ee
 we obtain the following canonical dual algebraic equation
\eb\label{PhiS}
\Phi (\sigma)^2 = 2\left[\sigma \tau(\sigma)\right]^2 ( \sigma - H_2) = H_1.
\ee
\section{Complete Solutions}
From the canonical duality theory \cite{gao-book00}, we have the following result:
\begin{thm}\label{thm1}
Let $\bar\sigma \in \real$ be a solution of the dual algebraic equation (\ref{PhiS}). Then, the vector ${\bf \bar x} = {\bf x}(\bar\sigma)$ is a critical point of $P_n({\bf x})$ and
\[
P_n({\bf \barx})=\Xi({\bf \barx}, \bar\sigma)=P^d(\bar\sigma).
\]
Conversely, every critical point ${\bf \bar x}$ of $P_n({\bf x})$ can be put in form (\ref{xrels}) for some solution $\bar \sigma \in \real$ to (\ref{PhiS}), again with
\[
P_n({\bf \barx})=\Xi({\bf \barx}, \bar\sigma)=P^d(\bar\sigma).
\]
\end{thm}
{\em Proof}.
By the definition of the canonical dual function, we know that
\[
\Xi({\bf \barx}, \bar\sigma)=P^d(\bar\sigma).
\]
Since $\bar\sigma$ is a solution to $\Phi(\bar\sigma)^2=H_1$, we can combine (\ref{xrels}) and (\ref{PhiS}) to get
\eb\label{srelx}
\bar\sigma= a_1 \bar y_1 + b_1,
\ee
where
\[
\bar y_1 = y_1({\bf \bar x}).
\]
Plugging this into (\ref{xrels}) gives the criticality condition
\eb\label{xcrit}
(a_2 \bar y_2+ b_2)(a_1 \bar y_1 + b_1)(a_0 {\bf \bar x + b_0})-{\bf h}=0,
\ee
where
\[
\bar y_2 = y_2(\bar y_1),
\]
and thus ${\bf \bar x}$ is a critical point of $P_n({\bf x})$. Furthermore,
\[
P_n({\bf \barx})=\Xi({\bf \barx}, \bar\sigma)=P^d(\bar\sigma)
\]
by construction.

Conversely, if ${\bf \bar x}$ is a critical point of $P_n({\bf x})$, we know that (\ref{xcrit}) is satisfied.
Plugging in ${\bf \barx}= {\bf x}(\bar\sigma)$ and simplifying gives us (\ref{PhiS}), and thus we know that $\bar\sigma$ is a solution to the algebraic equation. Additionally, the condition ${\bf \bar x}= {\bf x}(\bar\sigma)$ leads to $\Xi({\bf \barx}, \bar\sigma)= P^d(\bar\sigma)$, and the criticality condition combined with (\ref{xrels}) leads to (\ref{srelx}) and thus $P_n({\bf \barx})=\Xi({\bf \barx}, \bar\sigma)=P^d(\bar\sigma)$. This completes the proof. \hfill $\Box$ \\

\begin{remark}
From (\ref{dPds}), it is clear that $P^d(\sigma)$ has two extra critical points at $\pm \sqrt{\frac{H_3}{3}}$. However, from Theorem 1, we know that only ${\bf\bar x}= {\bf x}(\bar\sigma)$, where $\bar\sigma$ solve (\ref{PhiS}), are critical points of $P_n({\bf x})$. Even though $\pm \sqrt{\frac{H_3}{3}}$ are critical points of $P^d(\sigma)$, ${\bf x}(\pm \sqrt{\frac{H_3}{3}})$ may not be critical points of $P_n(\bx)$.
 These critical points were first noticed in \cite{gao-jogo06} but their origin was unexplained. Additionally, it is also clear from (\ref{dPds}) that solutions $\bar\sigma$ to (\ref{PhiS}) are not necessarily critical points of $P^d(\sigma)$ when ${\bf h}={\bf 0}$. When ${\bf h}\ne {\bf 0}$, all solutions to (\ref{PhiS}) are also critical points of $P^d(\sigma)$.
\end{remark}

From this remark we know that  the extra critical points   $\sigma= \pm\sqrt{\frac{H_3}{3}}$ of  $P^d(\sigma)$
do not correspond to critical points of $P_n({\bf x})$. Instead, it can be inferred from
\eb\label{dXids}
\frac{\partial\Xi ({\bf x}, \sigma)}{\partial\sigma}= \frac{a_2 (3 \sigma^2 - H_3)}{2 a_1}\left(y_1({\bf x}) - \frac{\sigma - b_1}{a_1}\right)= \frac{d[\sigma \tau(\sigma)]}{d\sigma}\left(y_1({\bf x}) - \frac{\sigma - b_1}{a_1}\right)
\ee
and (\ref{XiSimp}) that these two critical points are caused by the introduction of the second dual variable $s_2$ and its subsequent elimination by substitution, since the term $\frac{d [\sigma \tau(\sigma) ]}{d\sigma}$ gives precisely the extra critical points $\pm\sqrt{\frac{H_3}{3}}$. We can define the term ``corresponding critical point" to denote a solution $\bar \sigma$ of the algebraic equation (\ref{PhiS}): from Theorem \ref{thm1}, we know that all corresponding critical points $\bar\sigma$ are part of an (${\bf \barx}$, $\bar\sigma$) pair such that ${\bf \bar x}= {\bf x}(\bar\sigma)$ is a critical point of $P({\bf x})$ and $P^d(\bar \sigma)=P({\bf \bar x})$.

We also know that every critical point ${\bf \barx}$ of $P_n({\bf x})$ is part of an (${\bf \barx}$, $\bar\sigma$) pair such that the above conditions hold.

\subsection{Complete Solutions for the case ${\bf h}={\bf 0}$ }
From (\ref{dPds}) it is clear that $\bar\sigma$ is a critical point of $P^d(\sigma)$ and $({\bf \barx}, \bar\sigma)$ is a critical point of $\Xi({\bf x}, \sigma)$ if ${\bf h} \ne {\bf 0}$. If ${\bf h} = {\bf 0}$, then $\bar\sigma$ is a critical point of $P^d(\sigma)$ and $({\bf \barx}, \bar\sigma)$ is a critical point of $\Xi({\bf x}, \sigma)$ iff $\bar\sigma = H_2$. The other solutions to (\ref{PhiS}), $\bar\sigma = 0$ and $\bar\sigma^2 = H_3$, still correspond to critical points ${\bf \barx}$ of $P({\bf x})$ but are not themselves critical points of $P^d(\sigma)$.  Thus, we have:

\begin{thm}\label{h0crit}
Suppose that ${\bf \barx}$ is a critical point of $P_n({\bf x})$ and that ${\bf h}= {\bf 0}$. Then, either ${\bf \barx}$ is a solution to $a_1 \bary_1 + b_1=0$ and
\[
P_n({\bf\barx})= H_4 + \frac{a_2 H_3^2}{8 a_1^2},
\]
$a_1 \bary_1 + b_1 = H_2$ and
\[
P_n({\bf\barx})= H_4 + \frac{a_2 ( H_2^2 - H_3 )^2}{8 a_1^2},
\]
or $(a_1 \bary_1 + b_1)^2 = H_3$ and
\[
P_n({\bf\barx})= H_4.
\]
Furthermore, any ${\bf\barx}$ that satisfies $(a_1 \bary_1 + b_1)^2 = H_3$ is a global minimizer of $P_n({\bf x})$.
\end{thm}
{\em Proof}. From Theorem \ref{thm1}, we know that $P_n({\bf \barx})=\Xi({\bf \barx}, \bar\sigma)=P^d(\bar\sigma)$, where $\bar\sigma = a_1 \bary_1 + b_1$. Additionally, $\bar\sigma$ is a solution to (\ref{PhiS}), so we know that at least one of the following must be true:
\begin{eqnarray*}
a_1 \bary_1 + b_1 &=& 0\\
a_1 \bary_1 + b_1 &=& H_2\\
(a_1 \bary_1 + b_1)^2 &=& H_3.
\end{eqnarray*}
Furthermore,
\begin{eqnarray*}
P_n({\bf \barx})= P^d(0) &=& H_4 + \frac{a_2 H_3^2}{8 a_1^2}\\
P_n({\bf \barx})= P^d(H_2) &=& H_4 + \frac{a_2 (H_2^2 - H_3)^2}{8 a_1^2}\\
P_n({\bf \barx})= P^d(\pm \sqrt{H_3}) &=& H_4
\end{eqnarray*}
in each of those cases, respectively. Thus, it is clear that global minimizers ${\bf \barx}$ of $P_n({\bf x})$ satisfy the condition $(a_1 \bary_1 + b_1)^2 = H_3$. Additionally, since $H_2 \le \bar\sigma$, we also know that $\bar\sigma = 0$ only if $H_2 \le 0$ and that $\bar\sigma^2 = H_3$ only if $H_2^2 \le H_3$. This completes the proof. \hfill $\Box$ \\

\section{Global Extremality Conditions}
In order to identify both global and local extrema among all critical points, sufficient conditions are fundamentally important in global optimization.
 It can be seen from (\ref{srelx}) or (\ref{PhiS}) that $ H_2 \le \bar\sigma$ so we can assert that $\bar\sigma \in \calS_a$, where
\[
\calS_a = \{\sigma \in \real | \; H_2 \le \sigma\}.
\]
Additionally, if we define a region
\eb
\calS_a^+= \{ \sigma \in \calS_a | \Re(\sqrt{H_3}) < \sigma \},
\ee
where $\Re(z)$ is the real part of complex number $z$, the dual problem is
\eb
(\calP^d): \;\max \left\{P^d(\sigma) \; |\;\; \sigma \in \calS_a^+ \right\}.
\ee
and we obtain the following theorem.
\begin{thm}\label{globmin}
Suppose that $\bar\sigma$ is a critical point of $P^d(\sigma)$. If $\bar\sigma \in \calS_a^+$, then $\bar\sigma$ is a global maximizer of $P^d(\sigma)$ on $\calS_a^+$, ${\bf \barx}= {\bf x}(\bar\sigma)$ is a global minimizer of $P_n({\bf x})$, and
\[
P_n({\bf \barx})= \min_{{\bf x} \in \real^n} P_n({\bf x}) = \max_{\sigma \in \calS_a^+} P^d(\sigma) = P^d(\bar\sigma).
\]
\end{thm}
{\em Proof}. On the region $\sigma \in \calS_a^+$, $0 < \min \{ \frac{d[\sigma\tau(\sigma)]}{d\sigma}, \sigma, \tau(\sigma) \}$. Thus, we know from (\ref{dXids}) that
\[
P_n({\bf x})= \max_{\sigma \in \calS_a^+}\Xi({\bf x}, \sigma),
\]
and we can see from (\ref{dPds}) that $P^d(\sigma)$ is concave on $\sigma \in \calS_a^+$, thus $\bar\sigma$ is a global maximizer of $P^d(\sigma)$ on $\sigma \in \calS_a^+$. It is clear that the total complementary function
\[
\Xi({\bf x}, \sigma)= \Lam_1({\bf x}) \sigma \tau(\sigma) - U_1^*(\sigma) \tau(\sigma) - U_2^*(\tau(\sigma)) - {\bf h}^T {\bf x}
\]
is convex in ${\bf x} \in \real^n$, so we have
\begin{eqnarray*}
P^d(\bar\sigma) &=& \max_{\sigma \in \calS_a^+} P^d(\sigma)  =  \max_{\sigma \in \calS_a^+} \min_{{\bf x} \in \real^n} \Xi({\bf x}, \sigma) \\
&=& \min_{{\bf x} \in \real^n} \max_{\sigma \in \calS_a^+} \Xi({\bf x}, \sigma)= \min_{{\bf x} \in \real} P_n({\bf x}) = P_n({\bf \barx}).
\end{eqnarray*}
This completes the proof. \hfill $\Box$ \\

From (\ref{dPds}), it is clear that $P^d(\sigma)$ only has a critical point on $\calS_a^+$ at a solution to (\ref{PhiS}). Since
\[
\Phi(\sigma)^2= 2\left[\frac{a_2 \sigma (\sigma^2-H_3)}{2 a_1}\right]^2(\sigma - H_2),
\]
it is equally clear that $\Phi(\sigma)^2$ is monotonically increasing on $\calS_a^+$ and that $\Phi(\Re(\sqrt{H_3}))^2=0$. Thus, $P^d(\sigma)$ is concave on $\calS_a^+$ and, if ${\bf h} \ne {\bf 0}$, there is a guaranteed unique corresponding critical point $\bar\sigma$ on $\calS_a^+$. From Theorem \ref{h0crit}, we know that if ${\bf h}={\bf 0}$, there is no unique global minimizer of $P_n({\bf x})$. Instead, all solutions to $a_1 y_1({\bf x})+b_1=\pm\sqrt{H_3}$ are global minimizers of $P_n({\bf x})$.

\section{One-Dimensional Problem}
In the case case that $n=1$, both $h$ and $b_0$ are given constants and $(\calP_n)$ is presented as an extremum problem
\eb
(\calP)\;\;\;\;\ext \left\{P(x) = V(x) - h x
|\; x \in \real \right\}, \label{eq-8poly}
\ee
the notation $\ext \{ * \}$ stands for finding all extremum solutions. From Theorem \ref{thm1}, and the canonical duality theory, we know that $(\calP)$ corresponds to a dual problem
\eb
(\calP^d): \;\ext \left\{\calP^d(\sigma) \; |\;\; \sigma \in \real\right\}.
\ee
There is a natural dichotomy of problems of this type: the addition of the linear term $h x$ in $P(x)$ breaks the inherent symmetry and ensures one unique global minimizer of $P(x)$. Theorem \ref{h0crit} can be used to solve $(\calP)$ in the $h=0$ case.
\subsection{Local Extrema for $h \ne 0$}
In the case where $h \ne 0$, we have the following theorem.
\begin{thm}\label{thm2}
Suppose that $ h \ne 0$. The corresponding critical points $\bar\sigma$ of $P^d(\sigma)$, if they exist, lie in the regions defined by
\begin{eqnarray*}
\calS_a^- &=& \{ \sigma \in \calS_a |  H_2< \sigma < \Re(-\sqrt{H_3})\}\\
\calS_1 &=& \{ \sigma \in \calS_a | \Re(-\sqrt{H_3}) < \sigma < 0\}\\
\calS_2 &=& \{ \sigma \in \calS_a | 0 < \sigma < \Re(\sqrt{H_3})\}\\
\calS_a^+ &=& \{ \sigma \in \calS_a | \Re(\sqrt{H_3}) < \sigma \},
\end{eqnarray*}
where $\Re(z)$ is the real part of complex number $z$.
Furthermore, each of the regions $\calS_a^-$, $\calS_1$, and $\calS_2$, if non-empty, contains exactly one critical point, a maximizer, of $\Phi(\sigma)^2$. Finally, $\Phi(\sigma)^2$ is convex and monotone increasing on $\calS_a^+$.
\end{thm}
{\em Proof}. Because corresponding critical points $\bar\sigma$ of $P^d(\sigma)$ satisfy (\ref{PhiS}), it is clear that $\bar\sigma \in \calS_a^- \cup \calS_1 \cup \calS_2 \cup \calS_a^+$ when $h \ne 0$, since $\bar\sigma \in \calS_a$ and
\[
\Phi(\bar\sigma)^2 = H_1 =\frac{a_1 h^2}{a_0}\ne 0.
\]
Thus $\bar\sigma \ne H_2$ and $\bar\sigma \bar\tau \ne 0$.

Because $\Phi(\bar\sigma)^2 \ge 0$ and $\Phi(\sigma)^2 = 0$ at $\sigma = H_2$ and at $\sigma \tau(\sigma) = 0$, it follows that
each non-empty region $\calS_a^-$, $\calS_1$, and $\calS_2$ contains at least one maximizer of $\Phi(\sigma)^2$. Thus, if it can be shown that $\calS_a^-$, $\calS_1$, or $\calS_2$ contains exactly one critical point of $\Phi(\sigma)^2$, that critical point must be a maximizer of $\Phi(\sigma)^2$. Since
\begin{eqnarray*}
\frac{d[\Phi(\sigma)^2]}{d\sigma} &=& 2 \sigma \tau(\sigma)\left[\sigma \tau(\sigma) + 2(\sigma - H_2)\frac{d[\sigma \tau(\sigma)]}{d\sigma}\right]\\
&=& \frac{a_2}{a_1}\sigma \tau(\sigma) (7 \sigma^3 - 6 H_2 \sigma^2 - 3 H_3 \sigma + 2 H_2 H_3)\\
&=& \frac{a_2}{a_1}\sigma \tau(\sigma) Q(\sigma),
\end{eqnarray*}
with
\begin{eqnarray*}
Q(\sigma)&=& 7 \sigma^3 - 6 H_2 \sigma^2 - 3 H_3 \sigma + 2 H_2 H_3 \\
\frac{dQ(\sigma)}{d\sigma} = Q'(\sigma)&=& 3 (7 \sigma^2 - 4 H_2 \sigma - H_3),
\end{eqnarray*}
it follows that critical points of $\Phi(\sigma)^2$ occur in $\calS_a^-$, $\calS_1$, and $\calS_2$ only when $Q(\sigma)=0$ and that critical points of $Q(\sigma)$ occur at
\eb\label{spm}
\sigma^\pm= \frac{4 H_2 \pm \sqrt{16 H_2^2 + 28 H_3}}{14} = \frac{2 H_2 \pm \sqrt{4 H_2^2 + 7 H_3}}{7}.
\ee

If $\calS_a^- \ne \emptyset$, then $H_2 < \Re(-\sqrt{H_3}) \le 0$. When $H_3 = 0$, it is clear that $\Phi(\sigma)^2$ has exactly one critical point on $\calS_a^-$ at the point $\sigma= \frac{6 H_2}{7}$. When $H_3 > 0$, it is equally clear that $\calS_1$ and $\calS_2$ are non-empty, and $\calS_a^-$, $\calS_1$ and $\calS_2$ each contains at least one maximizer of $\Phi(\sigma)^2$ since we assume that $\calS_a^- \ne \emptyset$. Because $Q(\sigma)$ has at most three roots, it follows that $\calS_a^-$, $\calS_1$, and $\calS_2$ each have exactly one critical point, a local maximizer, of $\Phi(\sigma)^2$. When $H_3 < 0$, we know from (\ref{spm}) that $\frac{4 H_2}{7} < \sigma^\pm < 0$ and, since the expressions $\sigma^2(7 \sigma - 6 H_2)$ and $ H_3(2H_2 - 3\sigma)$ are both positive on this range of values, we know that $(\sigma^\pm)^2(7 \sigma^\pm - 6 H_2) +  H_3(2H_2 - 3\sigma^\pm)=Q(\sigma^\pm)>0$. Consequently, $\Phi(\sigma)^2$ will have exactly one critical point on $\calS_a^-$ if
\[
Q(H_2)Q\left(\Re(-\sqrt{H_3})\right) =  Q(H_2)Q(0)< 0.
\]
Calculating $Q(\sigma)$ at these points gives
\begin{eqnarray*}
Q(H_2) =H_2(H_2^2 - H_3) &<& 0\\
Q(0) =2 H_2 H_3 &>& 0,
\end{eqnarray*}
so $\calS_a^-$ contains exactly one critical point of $\Phi(\sigma)^2$ when $\calS_a^- \ne \emptyset$.

Similarly, when $\calS_1 \ne \emptyset$,
\begin{eqnarray*}
H_2 &<& 0 \\
H_3 &>& 0 \\
\calS_1 &=& \{\sigma \in \real | \; max \{H_2, - \sqrt{H_3} \} < \sigma < 0\}.
\end{eqnarray*}
Since it has already been shown that $\calS_a^-$, $\calS_1$, and $\calS_2$ each contains exactly one critical point of $\Phi(\sigma)^2$ when none of them are empty, and since $\calS_2 \ne \emptyset$ when $\calS_1 \ne \emptyset$, we can focus on the case where $\calS_a^- = \emptyset$. This implies that $- \sqrt{H_3} \le H_2$. From (\ref{spm}), we know that $\sigma^+>0 \notin \calS_1$, so
\[
Q(H_2)Q(0) < 0
\]
is a sufficient condition to prove that $\Phi(\sigma)^2$ has exactly one critical point on $\calS_1$.
Since
\begin{eqnarray*}
Q(H_2) =  H_2 (H_2^2- H_3) &\ge& 0\\
Q(0) = 2 H_2 H_3 &<& 0,
\end{eqnarray*}
it only remains to be shown that $\Phi(\sigma)^2$ has only one critical point on $\calS_1$ when $Q( H_2) = 0$.
In this case,
\[
Q'(\sigma)|_{\sigma= H_2} = 6 H_2 ^2 > 0,
\]
 so we know that $Q(\sigma) > 0$ for some $\sigma \in \calS_1$. Because $Q(\sigma)$ is continuous and has at most one critical point on $\calS_1$, it follows that $\Phi(\sigma)^2$ has only one critical point on $\calS_1$ when $\calS_1 \ne \emptyset$.

Thirdly, when $\calS_2 \ne \emptyset$,
\begin{eqnarray*}
H_3 &>& 0 \\
H_2 &<& \sqrt{H_3}\\
\calS_2 &=& \{\sigma \in \real | \; \max \{ H_2, 0\} < \sigma < \sqrt{ H_3 }\}.
\end{eqnarray*}
When $H_2 \le 0$, it is clear from (\ref{spm}) that $Q(\sigma)$ has at most one critical point on $\calS_2$, so
\[
Q(0) Q(\sqrt{H_3}) < 0
\]
is a sufficient condition to prove that $\Phi(\sigma)^2$ has exactly one critical point on $\calS_2$. Since
\begin{eqnarray*}
Q(0) =2 H_2 H_3  &\le& 0\\
Q(\sqrt{ H_3 }) = 4 H_3 ( \sqrt{H_3} - H_2) &>& 0
\end{eqnarray*}
and $Q'(\sigma)|_{\sigma=0} = -3 H_3 < 0$, there must be exactly one critical point of $\Phi(\sigma)^2$ on $\calS_2$ when $H_2 \le 0$. When $H_2>0$, there is at most one critical point of $Q(\sigma)$ in $\calS_2$ since $H_3 > 0$ and $\sigma^- < 0$. Thus, it is sufficient that
\[
Q(H_2)Q(\sqrt{ H_3})<0
\]
holds in order for $\Phi(\sigma)^2$ to have exactly one critical point on $\calS_2$. Because
\begin{eqnarray*}
Q(H_2) = H_2(H_2^2 - H_3) &<& 0\\
Q(\sqrt{H_3}) = 4 H_3 ( \sqrt{H_3} - H_2) &>& 0,
\end{eqnarray*}
$\Phi(\sigma)^2$ has exactly one critical point on $\calS_2$ when $\calS_2 \ne \emptyset$.

Finally, $\Phi(\sigma)^2$ is convex on $\calS_a^+$ because $\sigma>0$, $\tau(\sigma)>0$, $\frac{d[\sigma \tau(\sigma)]}{d\sigma}>0$, $\frac{d^2[\sigma \tau(\sigma)]}{d\sigma^2}>0$, and $H_2 < \sigma$ for $\sigma \in \calS_a^+$. Thus, all terms in
\[
\frac{d^2[\Phi(\sigma)^2]}{d\sigma^2}= 2 \left( 4 \sigma \tau(\sigma) \frac{d[\sigma \tau(\sigma)]}{d\sigma} +
2 (\sigma - H_2) \left[ \sigma \tau(\sigma) \frac{d^2[\sigma \tau(\sigma)]}{d\sigma^2} + \left( \frac{d[\sigma \tau(\sigma)]}{d\sigma}\right)^2\right] \right)
\]
are positive. By the same reasoning, $\frac{d[\Phi(\sigma)^2]}{d\sigma}>0$ and $\Phi(\sigma)^2$ is also monotone increasing on $\calS_a^+$.  This completes the proof.
\[
\]

We can label the maximizers of $\Phi(\sigma)^2$ on $\calS_a^-$, $\calS_1$, and $\calS_2$ as $\sigma^\flat$, $\sigma^\natural$, and $\sigma^\sharp$, respectively, and define
\begin{eqnarray*}
\calS_{a+}^- &=& \{ \sigma \in \calS_a^- | \; \sigma< \sigma^\flat\} \\
\calS_{a-}^- &=& \{ \sigma \in \calS_a^- | \; \sigma> \sigma^\flat\} \\
\calS_{1+} &=& \{ \sigma \in \calS_b^- | \; \sigma< \sigma^\natural\} \\
\calS_{1-} &=& \{ \sigma \in \calS_b^- | \; \sigma> \sigma^\natural\} \\
\calS_{2+} &=& \{ \sigma \in \calS_b^+ | \; \sigma< \sigma^\sharp\} \\
\calS_{2-} &=& \{ \sigma \in \calS_b^+ | \; \sigma> \sigma^\sharp\}.
\end{eqnarray*}
From Theorem \ref{thm2}, we know that $\frac{d\Phi(\sigma)^2}{d\sigma}>0$ on $\calS_{a+}^-$, $\calS_{1+}$, and $\calS_{2+}$ and that $\frac{d\Phi(\sigma)^2}{d\sigma}<0$ on $\calS_{a-}^-$, $\calS_{1-}$, and $\calS_{2-}$. If we set $R(\sigma)$ to the expression obtained when (\ref{rela}) and (\ref{xrels}) are substituted into $\frac{d^2P(x)}{dx^2}$
\begin{eqnarray*}
R(\sigma) =   \left\{ \frac{d^2 P(x)}{dx^2}| \; x= \frac{1}{a_0}\left(\frac{h}{\sigma \tau(\sigma)} - b_0\right) \right\} = \frac{a_0 a_2}{2 a_1}(7 \sigma^3 - 6 H_2 \sigma^2 - 3 H_3 \sigma + 2 H_2 H_3)= \frac{a_0 a_2}{2 a_1} Q(\sigma),
\end{eqnarray*}
it is clear that $Q(\sigma)R(\sigma) >0$ when $Q(\sigma) \ne 0$ and, since $R(\bar\sigma)= \frac{d^2 P(\barx)}{dx^2}$, that $Q(\bar\sigma)\frac{d^2 P(\barx)}{dx^2}>0$ when $Q(\bar\sigma) \ne 0$ as well. When $Q(\bar\sigma)=0$, then $\frac{d^2P(\barx)}{dx^2}=0$. Thus, we have the following theorem.

\begin{thm}\label{thm3}
Suppose that  $h \ne 0$. Then, each of the regions $\calS_{a\pm}^-$, $\calS_{i\pm} \; (i= 1, 2)$ contains at most one corresponding critical point $\bar\sigma$ of $P^d(\sigma)$, $\calS_a^+$ contains exactly one corresponding critical point of $P^d(\sigma)$, and
$\bar\sigma \in \calS_{a\pm}^- \Leftrightarrow \bar\omega \in \calS_{a\mp}^-$, $\bar\sigma \in \calS_{i\pm} \Leftrightarrow \bar\omega \in \calS_{i\mp}$, where $\bar\omega$ is another corresponding critical point of $P^d(\sigma)$. Finally, $\bar\sigma \in \calS_{a+}^- \cup \calS_{1-} \cup \calS_{2+}$ if and only if $\barx= x(\bar\sigma)$ is a local maximizer of $P(\barx)$, $\bar\sigma \in \calS_{a-}^- \cup \calS_{1+} \cup \calS_{2-} \cup \calS_a^+$ if and only if $\barx$ is a local minimizer of $P(\barx)$, and $\bar\sigma \in \sigma^\flat \cup \sigma^\natural \cup \sigma^\sharp$ if and only if $\barx$ is an inflection point of $P(x)$.
\end{thm}
{\em Proof}. Since we know that $\frac{d\Phi(\sigma)^2}{d\sigma}>0$ on $\calS_{a+}^- \cup \calS_{1+} \cup \calS_{2+} \cup \calS_a^+$, and that $\frac{d\Phi(\sigma)^2}{d\sigma}<0$ on $\calS_{a-}^- \cup \calS_{1-} \cup \calS_{2-}$ from Theorem \ref{thm2}, we also know that each of these regions can have no more than one value at which $\Phi(\sigma)^2 = H_1$ because $\Phi(\sigma)^2$ is continuous over $\sigma \in \real$. Since
\[
x(\sigma) = \frac{1}{a_0}\left(\frac{h}{\sigma \tau(\sigma)} - b_0\right)
\]
is a critical point of $P(x)$ if and only if $\Phi(\sigma)^2=H_1$, it follows that there can be at most one $\bar\sigma$ on each region. Also, since $\Phi(\sqrt{H_3})^2=0 < H_1$ and  $\Phi(\sigma)^2$ is monotonically increasing on $\calS_a^+$, there is exactly one $\bar\sigma \in \calS_a^+$. Because $\Phi(H_2)^2= 0 < H_1$, $\sigma^\flat > H_1$ iff $\calS_{a+}^-$ contains one corresponding critical point. Since $\Phi(-\sqrt{ H_3})^2 = 0 < H_1$, $\calS_{a-}^-$ contains a corresponding critical point iff $\sigma^\flat>H_1$. Therefore, $\calS_{a+}^-$ contains a corresponding critical point $\bar \sigma$ iff $\calS_{a-}^-$ does as well.  Because $ \Phi(\pm \sqrt{H_3}) = \Phi(0)=0 < H_1$, the same logic ensures that $\calS_{i+} \; (i= 1, 2)$ contains a corresponding critical point iff $\calS_{i-}$ does.

On $\calS_a^-$ and $\calS_2$, $\sigma \tau(\sigma)<0$ so $Q(\sigma)<0$ on $\calS_{a+}^-  \cup \calS_{2+}$ and $Q(\sigma)>0$ on $\calS_{a-}^-  \cup \calS_{2-}$. By the same logic, $Q(\sigma)>0$ on $\calS_{1+}  \cup \calS_a^+$ and $Q(\sigma)<0$ on $\calS_{1-}$. Thus,
\begin{eqnarray*}
\calS_{a+}^-  \cup \calS_{1-} \cup \calS_{2+} &=& \{\sigma \in \calS_a | \; Q(\sigma) < 0, \sigma\tau(\sigma) \ne 0\} \\
\calS_{a-}^-  \cup \calS_{1+} \cup \calS_{2-} \cup \calS_a^+ &=& \{\sigma \in \calS_a | \; Q(\sigma) > 0, \sigma\tau(\sigma) \ne 0\}.
\end{eqnarray*}
Consequently,
\begin{eqnarray*}
\frac{d^2P(\barx)}{dx^2}<0 &\Leftrightarrow& \bar\sigma \in \calS_{a+}^-  \cup \calS_{1-} \cup \calS_{2+}\\
\frac{d^2P(\barx)}{dx^2}>0 &\Leftrightarrow& \bar\sigma \in \calS_{a-}^-  \cup \calS_{1+} \cup \calS_{2-} \cup \calS_a^+.
\end{eqnarray*}
Since $H_1 \ne 0$,
\[
\sigma^\flat \cup \sigma^\natural \cup \sigma^\sharp = \{ \sigma \in \calS_a^- \cup \calS_1 \cup \calS_2 \cup \calS_a^+ | \; Q(\sigma)=0 \},
\]
thus
\[
\frac{d^2P(\barx)}{dx^2}=0 \Leftrightarrow \bar\sigma \in \sigma^\flat \cup \sigma^\natural \cup \sigma^\sharp.
\]
This completes the proof. \hfill $\Box$ \\

\subsection{Classification of critical points}
In addition to knowing where critical points $\barx$ occur, it would also be useful to know how many critical points $P(x)$ has based on the particular values of the coefficients. In the case where $h=0$, this amounts to comparing $H_2$, $H_3$ and $0$: $P(x)$ has seven critical points when $H_2 <  \Re (-\sqrt{H_3}) <0$, five critical points when
$\Re (-\sqrt{H_3}) \le H_2 < 0$, three critical points when $0 \le H_2 < \Re ( \sqrt{H_3} )$ or $H_2 < 0 = \Re ( \sqrt{H_3})$, and one critical point when $0 \le \Re(\sqrt{H_3})<H_2$. In other words, $P(x)$ has seven critical points when $\calS_a^-$, $\calS_1$, $\calS_2$, and $\calS_a^+$ are all non-empty, five critical points when only $\calS_a^-$ is empty, three critical points when either $\calS_a^-$ or $\calS_2$ is the only non-empty region apart from $\calS_a^+$, and only one critical point when $\calS_a^-$, $\calS_1$, and $\calS_2$ are all empty.

In the case where $h \ne 0$, the number of critical points of $P(x)$ also depends on the value of $H_1$: if all other coefficients stay constant, $P(x)$ has a maximum number of critical points when $H_1=h=0$ and only loses critical points as $H_1$ increases in value. 

\section{Applications}
\subsection{Case n=1 }

\begin{figure}[h]
\begin{center}
\scalebox{.5}{\includegraphics{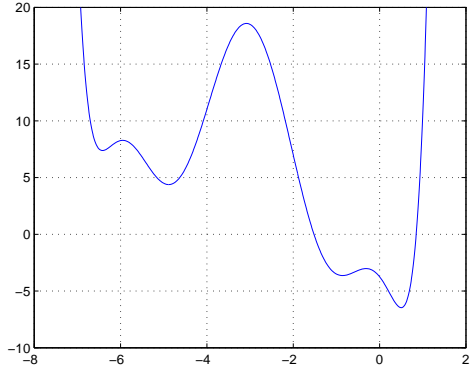}}
\caption{Graph of $P(x)$}\label{P1}
\end{center}
\end{figure}

Letting $a_i=1 \; (i= 0, 1, 2)$, $b_0=3$, $b_1=2$, $b_2=1$,
$c_0= -1.5$, $c_1=-1$, $c_2=-5$, and $h=2$, we have
\[
P (x) = \frac{1}{128} x^8 + \frac{3}{16} x^7 + \frac{55}{32} x^6 + \frac{117}{16} x^5 + \frac{851}{64} x^4 + \frac{69}{16} x^3 - \frac{249}{32} x^2 - \frac{77}{16} x - \frac{479}{128}.
\]
The graph of $P(x)$ is shown in Fig. \ref{P1}.
\begin{figure}[h]
\begin{center}
\scalebox{.5}{\includegraphics{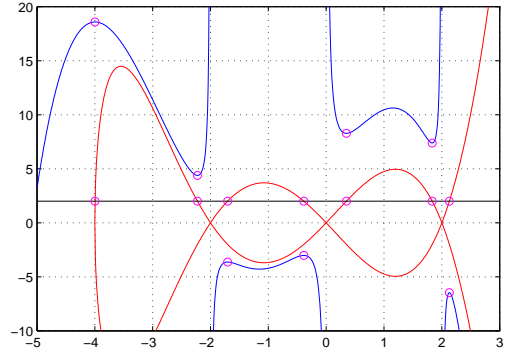}}
\caption{Graphs of $P^d (\sigma)$, $\Phi(\sigma)$, and $\sqrt{H_1}$}\label{Pd1}
\end{center}
\end{figure}

The canonical dual of $P(x)$ has the form
\[
P^d (\sigma) = \frac{1}{2} + \frac{(\sigma^2 - 4)^2}{8} - \frac{\sigma^2 (\sigma^2 - 4)^2(\sigma + 4) + 8}{2 \sigma (\sigma^2 - 4)}
\]
and the constants $H_1$, $H_2$, and $H_3$ have the values
\begin{eqnarray*}
H_1 &=& \frac{a_1 h^2}{a_0} = 4\\
H_2 &=& \frac{2 a_0 a_1 c_0 + 2 a_0 b_1 - a_1 b_0^2}{2 a_0} = -4\\
H_3 &=& -\frac{2 a_1 a_2 c_1 + 2 a_1 b_2 - a_2 b_1^2}{a_2} = 4.\\
\end{eqnarray*}
Its graph, in blue, is shown in Fig. \ref{Pd1} along with $\Phi(\sigma) = \pm \sigma \tau(\sigma) \sqrt{2(\sigma - H_2)}$ in red and $\sqrt{H_1} = h \sqrt{\frac{a_1}{a_0}}$ in black. Corresponding critical points $\bar\sigma$ are circled.
Since
\[
H_1 = 4 < \Phi(\sigma^\natural)^2 < \Phi(\sigma^\sharp)^2 <\Phi(\sigma^\flat)^2,
\]
there are seven corresponding critical points $\bar\sigma$ of $P^d(\sigma)$ and seven critical points $\barx$ of $P(x)$.
\begin{figure}[h]
\begin{center}
\scalebox{.5}{\includegraphics{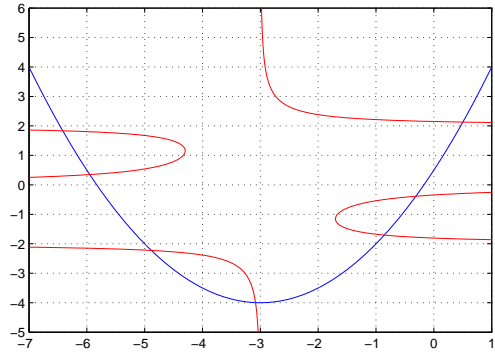}}
\caption{Graphs of $x(\sigma)$ (red) and $\sigma(x) = a_1 y_1(x) + b_1$ (blue) with $x$ as the abscissa and $\sigma$ as the ordinate.}\label{xs1}
\end{center}
\end{figure}

It can be seen from Fig. \ref{xs1} that the $(\barx, \bar\sigma)$ pairs are, in descending value of $\bar\sigma$,
\begin{eqnarray*}
(\barx_1, \bar\sigma_1) &=& (.05014, 2.1299),\\
(\barx_2, \bar\sigma_2) &=& (-6.4157, 1.8334),\\
(\barx_3, \bar\sigma_3) &=& (-5.9495, 0.3497),\\
(\barx_4, \bar\sigma_4) &=& (-0.3117, -0.3864),\\
(\barx_5, \bar\sigma_5) &=& (-0.8573, -1.7043),\\
(\barx_6, \bar\sigma_6) &=& (-4.8837, -2.2258),\\
(\barx_7, \bar\sigma_7) &=& (-3.0836, -3.9965).
\end{eqnarray*}

\begin{figure}[h]
\begin{center}\scalebox{.5}{\includegraphics{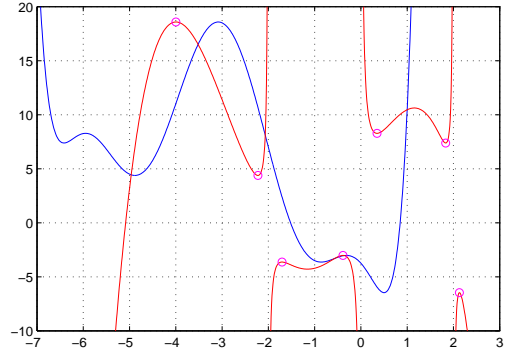}}
\caption{Graphs of $P(x)$ (blue) and $P^d(\sigma)$ (red) for $H_1 < \min\{\Phi (\sigma^\flat)^2, \Phi (\sigma^\natural)^2, \Phi (\sigma^\sharp)^2\}$}\label{PPd1}
\end{center}
\end{figure}

If we overlay $P(x)$ and $P^d(\sigma)$, as in Fig. \ref{PPd1}, we can clearly see that $\barx_1$ is the global minimizer of $P(x)$ on $x \in \real$ and that $\bar\sigma_1$ is the global maximizer of $P^d(\sigma)$ on $\sigma \in \calS_a^+$. Thus, $\barx_1$ is the solution to $(\calP_1)$ and $\bar\sigma_1$ is the solution to $(\calP^d)$ over $\sigma \in \calS_a^+$. Also, it can be seen both from Figs. \ref{xs1} and \ref{PPd1} that $P(\barx)=P^d(\bar\sigma)$ for every $(\barx, \bar\sigma)$ since  Fig. \ref{xs1} traces out the regions in the $x\sigma$-plane where $\Xi(x, \sigma)= P(x)$ and $\Xi(x, \sigma)=P^d(\sigma)$ in blue and red, respectively.

\begin{figure}[h]
\begin{center}\scalebox{.45}{\includegraphics{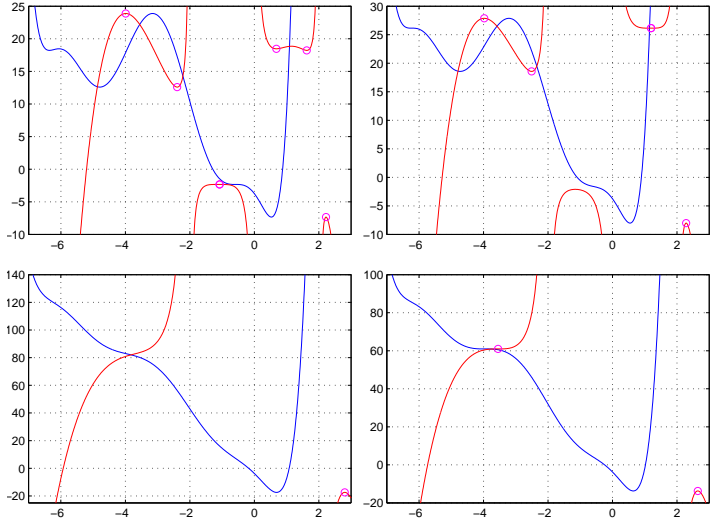}}
\caption{Clockwise from top left, $P(x)$ and $P^d(\sigma)$ for $\sqrt{H_1} = h = |\Phi (\sigma^\natural)| = 3.6978$, $h =|\Phi (\sigma^\sharp)| = 4.9535$, $h = |\Phi (\sigma^\flat)| = 14.4859$, and $h = 20$}\label{PPd1cases}
\end{center}
\end{figure}

Since the critical point of $\Phi(\sigma)^2$ on $\calS_a^i \; (i= 1, 2, 3)$ occurs at the same value of $\sigma$ that maximizes $|\Phi(\sigma)|$ on $\calS_a^i$, that is, maximums of $|\Phi(\sigma)|$ on $\calS_a^i \; (i= 1, 2, 3)$ occur at $\sigma^\flat$, $\sigma^\natural$, and $\sigma^\sharp$ respectively, we can see from Fig. \ref{PPd1cases} that $\bar\sigma \in \calS_a^i$ corresponds to an inflection point of $P(x)$ when $|\Phi(\bar\sigma)| \ge |\Phi(\sigma)|$ for all $\sigma \in \calS_a^i$ and that $\bar\sigma \notin \calS_a^i$ if
\[
\max_{\sigma \in \calS_a^i}|\Phi(\sigma)| < \sqrt{H_1}.
\]
As always, $\bar\sigma \in \calS_a^+$ corresponds to the global minimizer of $P(x)$.

\begin{figure}[h]
\begin{center}\scalebox{.5}{\includegraphics{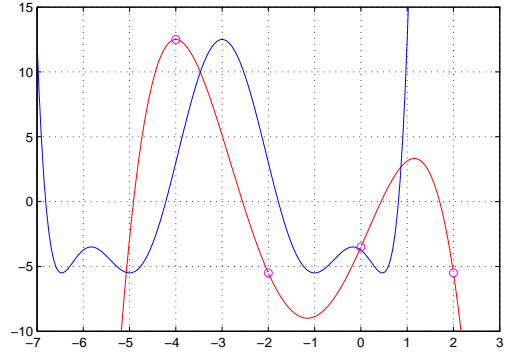}}
\caption{$P(x)$ and $P^d(s)$ for $h=0$}\label{PPd1a}
\end{center}
\end{figure}

When $h = 0$, the overlay of $P(x)$ in blue and $P^d(\sigma)$ in red is shown by Fig. \ref{PPd1a}. It can be seen that even though the circled points on $P^d(\sigma)$ corresponding to solutions of $\Phi(\sigma)^2=H_1$ are not critical points of $P^d(\sigma)$, they still correspond to critical points of $P(x)$.

\subsection{Case n=2}
\begin{figure}[h]
\begin{center}
\scalebox{.5}{\includegraphics{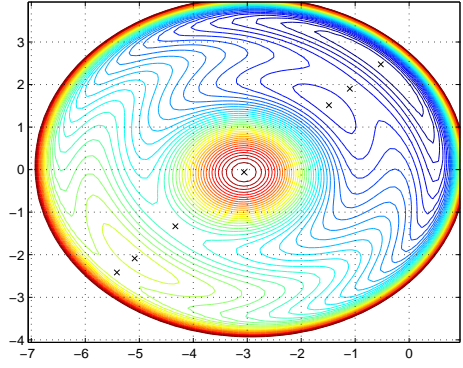}}
\caption{Contour of $P_2({\bf x})$. Warmer colors correspond to larger values of $P({\bf x})$.}\label{P2D}
\end{center}
\end{figure}

If we set $a_i=1 \; (i= 0, 1, 2)$, ${\bf b_0}=[3 \;0]^T$, $b_1=2$, $b_2=1$,
$c_0= -1.5$, $c_1=-1$, $c_2=-1$, and ${\bf h}=\sqrt{2} [1 \;1]^T$, we get
\begin{eqnarray*}
H_1 &=& \frac{a_1 \|{\bf h}\|^2}{a_0}= 4\\
H_2 &=& \frac{2 a_0 a_1 c_0 + 2 a_0 b_1 - a_1 \|{\bf b_0}\|^2}{2 a_0} = -4\\
H_3 &=& -\frac{2 a_1 a_2 c_1 + 2 a_1 b_2 - a_2 b_1^2}{a_2} = 4.\\
\end{eqnarray*}
as before; the canonical dual $P^d(\sigma)$ also has the same form, and is shown in Fig. \ref{Pd1}. A contour plot  of $P_2({\bf x})$ is shown in Fig. \ref{P2D}. Critical points are crossed.

The corresponding critical points $\bar\sigma$ are the same as in the previous example, but the $({\bf \barx}, \bar\sigma)$ pairs are now, in descending value of $\bar\sigma$,
\begin{eqnarray*}
({\bf \barx_1}, \bar\sigma_1) &=& ([-0.525, 2.475]^T, 2.1299),\\
({\bf \barx_2}, \bar\sigma_2) &=& ([-5.416, -2.416]^T, 1.8334),\\
({\bf \barx_3}, \bar\sigma_3) &=& ([-5.086, -2.086]^T, 0.3497),\\
({\bf \barx_4}, \bar\sigma_4) &=& ([-1.099, 1.901]^T, -0.3864),\\
({\bf \barx_5}, \bar\sigma_5) &=& ([-1.485, 1.515]^T, -1.7043),\\
({\bf \barx_6}, \bar\sigma_6) &=& ([-4.332, -1.332]^T, -2.2258),\\
({\bf \barx_7}, \bar\sigma_7) &=& ([-3.059, -0.059]^T, -3.9965).
\end{eqnarray*}

From Figs. \ref{Pd1} and \ref{P2D}, we can see that ${\bf \barx_1}$ is the global minimizer of $P_2({\bf x})$ on ${\bf x} \in \real^2$ and that $\bar\sigma_1$ is the global maximizer of $P^d(\sigma)$ on $\sigma \in \calS_a^+$. Thus, ${\bf \barx_1}$ is the solution to $(\calP_2)$ and $\bar\sigma_1$ is the solution to $(\calP^d)$ over $\sigma \in \calS_a^+$.

\section{Conclusions}
In this paper we obtain the exact solution to the 8th order polynomial minimization problem, which was seldom discussed before.
One advantage of using this way to find the critical points of  $P_n({\bf x})$ is that the largest corresponding critical point of $P^d(\sigma)$ corresponds to the global minimizer of $P_n({\bf x})$. Furthermore, we can see the efficiency of the canonical dual transformation method: it provides powerful primal-dual alternate approaches with zero duality gap and can reduce the number of dimensions in nonlinear programming.

\subsection*{Acknowledgements}
Special thanks to Professor David Y. Gao for advice, suggestions, and encouragement.
This research is partially supported by US Air Force
Office of Scientific Research under the grant AFOSR FA9550-10-1-0487.

\end{document}